\documentclass{elsart}

\usepackage{epsfig,caption,multirow,subfigure}

\begin {document}
\begin{frontmatter}
\title{Background Light in Potential Sites \\for the ANTARES\\
       Undersea Neutrino Telescope}

\author[10]{P.~Amram}, 
\author[4]{S.~Anvar}, 
\author[1]{E.~Aslanides}, 
\author[1]{J-J.~Aubert}, 
\author[4]{R.~Azoulay}, 
\author[1]{S.~Basa}, 
\author[5]{Y.~Benhammou}, 
\author[1]{F.~Bernard}, 
\author[1]{V.~Bertin}, 
\author[1]{M.~Billault}, 
% ...
. \\

\vspace{1cm}
\centerline{The ANTARES Collaboration}

\address[1]{Centre de Physique des Particules de Marseille (CPPM), (CNRS/IN2P3 - Universit\'e de la M\'editerran\'ee Aix-Marseille II), 163 Avenue de Luminy, Case 907, 13288 Marseille Cedex 09, France}
\address[2]{Centre de Physique Th\'eorique (CPT), (CNRS), 163 Avenue de Luminy, Case 907, 13288 Marseille Cedex 09, France}
\address[3]{Centre d'Oc\'eanologie de Marseille, (CNRS/INSU - Universit\'e de la M\'editerran\'ee), Station Marine d'Endoume-Luminy, Rue de la Batterie des Lions, 13007 Marseille, France}
\address[4]{DAPNIA/DSM, CEA/Saclay, 91191 Gif Sur Yvette Cedex, France}
\address[5]{Groupe de Recherches en Physique des Hautes Energies (GRPHE), (Universit\'e de Haute Alsace), 61 Rue Albert Camus, 68093 Mulhouse Cedex, France}
\address[6]{IFREMER, Centre de Toulon/La Seyne Sur Mer, Port Bregaillon, Chemin Jean-Marie Fritz, 83500 La Seyne Sur Mer, France IFREMER, Centre de Brest, 29280 Plouzan\'e, France}
\address[7]{Institut de Recherches Subatomiques (IReS), (CNRS/IN2P3 - Universit\'e Louis Pasteur), BP 28, 67037 Strasbourg Cedex 2, France}
\address[8]{Instituto de F\'{\i}sica Corpuscular (IFIC), CSIC - Universitat de Val\`encia, 46100 Burjassot, Valencia, Spain}
\address[9]{Laboratoire d'Astronomie Spatiale, Institut Gassendi pour la Recherche Astronomique en Provence (IGRAP), (CNRS/INSU - Universit\'e de Provence Aix-Marseille I), Les Trois Lucs, Traverse du Siphon, 13012 Marseille Cedex, France}
\address[10]{Observatoire de Marseille, Institut Gassendi pour la Recherche Astronomique en Provence (IGRAP), (CNRS/INSU - Universit\'e de Provence Aix-Marseille I), 2 Place Le Verrier, 13248 Marseille Cedex 4, France}
\address[11]{University of Oxford, Department of Physics, Nuclear and Astrophysics Laboratory, Keble Road, Oxford OX1 3RH, United Kingdom}
\address[12]{University of Sheffield, Department of Physics and Astronomy, Sheffield S3 7RH, United Kingdom}

\newpage
\begin{abstract}

The ANTARES collaboration has performed a series of {\em in situ}
measurements to study the background light for a planned undersea
neutrino telescope. Such background can be caused by $^{40}$K decays
or by biological activity. We report on measurements at two sites in
the Mediterranean Sea at depths of 2400~m and 2700~m, respectively.
Three photomultiplier tubes were used to measure single counting rates
and coincidence rates for pairs of tubes at various distances. The
background rate is seen to consist of three components: a constant
rate due to $^{40}$K decays, a continuum rate that varies on a time
scale of several hours simultaneously over distances up to at least
40~m, and random bursts a few seconds long that are only correlated in
time over distances of the order of a meter. A trigger requiring
coincidences between nearby photomultiplier tubes should reduce the
trigger rate for a neutrino telescope to a manageable level with only
a small loss in efficiency.
\end{abstract}

\end{frontmatter}

\vskip 10cm
\begin{tabbing}
{\em Correspondence to:} \= N. Palanque-Delabrouille, DAPNIA/DSM, CEA/Saclay,\\
\> 91191 Gif Sur Yvette Cedex, France. \\
\> Nathalie.Delabrouille@cea.fr
\end{tabbing}

\newpage
\normalsize
\section{INTRODUCTION}

\begin{figure}[h] \begin{center} 
  \epsfig{file=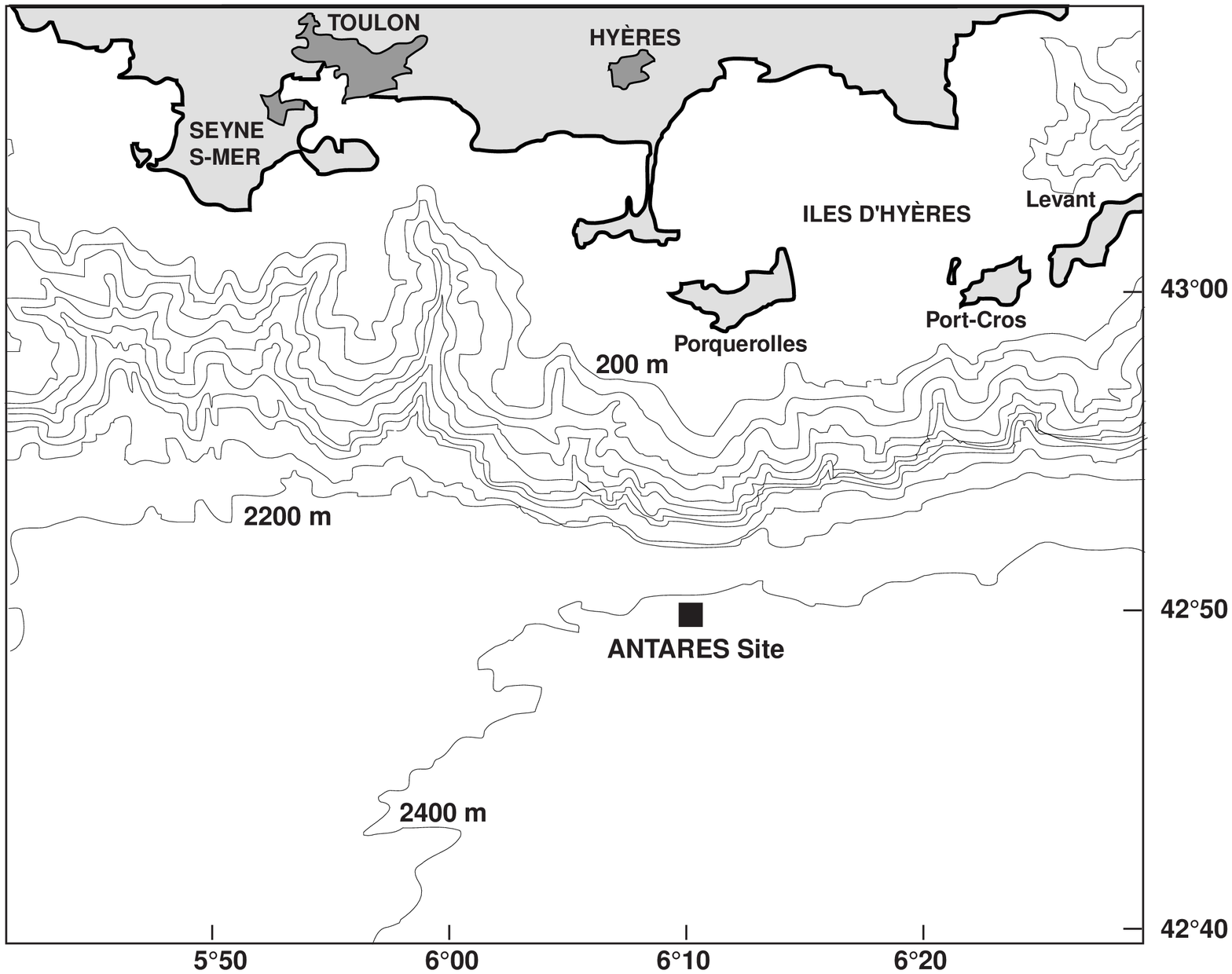,width=.8\textwidth} \caption{Location
  of the ANTARES site, near the Mediterranean French
  coast.} \label{site} \end{center}
\end{figure}

The ANTARES\footnote{Astronomy with a Neutrino Telescope and Abyss
environmental RESearch.\\ANTARES WEB site:
http://antares.in2p3.fr/antares/antares.html} project \cite{proposal}
leads to the deployment of an undersea neutrino telescope (this
concept was first proposed in~\cite{markov}) with an area of
1/10~km$^2$ in a Mediterranean site 20 nautical miles off the coast
from Toulon (France) at a depth of 2400~m. The location of the site is
shown on figure~\ref{site}.  The ANTARES detector is aimed at the
observation of neutrinos from astrophysical sources, an indirect
detection of dark matter and the study of neutrino oscillations. An
array of photomultiplier tubes detects the Cherenkov light emitted in
the sea water from the muons produced by the neutrinos in the
surrounding medium. At high energy, the muon is emitted in a direction
close to that of the parent neutrino --- $\theta_{\nu\mu} \simeq
0.7^\circ / E^{0.6}(TeV)$ --- and its direction is derived from the
arrival times of the Cherenkov light wave front on at least 5~optical
modules (where each optical module contains a single photomultiplier
tube).  The detector will consist of about 1000 optical modules on
several vertical strings read out via a single optical cable to shore.

As part of this project, we have developed measuring systems to
characterize candidate detector sites for a proposed high energy
neutrino facility. The selection criteria include the optical properties
of the sea water, the depth of the site, its distance to the shore and the
availability of technical support.

Autonomous systems have been designed to measure {\it in situ} the
background light, the fouling of glass spheres, and the light
attenuation. Surveys with time ranges of the order of the year were
carried out in order to study the seasonal variations of the sea
properties.

Background light signals are expected from $^{40}$K
decays\footnote{$\; ^{40}\rm K$ undergoes $\beta$-decay (89.3\%) and
electron capture (10.7\%).} and possibly from biological activity. A
knowledge of the background light behavior on site is of prime
importance as it puts constraints on the trigger logic and
electronics as well as on the mechanical layout of the optical
modules.  This paper reports on a series of tests dedicated to the
measurement of this background which were performed on two sites in
the Mediterranean sea.


\section{DESCRIPTION OF THE MEASUREMENT}


\subsection{Experimental set-up}

It is necessary to estimate how a detector will be affected by background
light sources, through the measure of counting rates, coincidence
rates, and their correlation at various distances. A set-up has
been devised which can be used to study the time dependence of
the background as well as its spatial extension and its correlation
with undersea current. It was immersed for durations ranging from
hours to months.

The measurements have been performed using optical modules with the
same design as the ones which will be used for the final detector,
except for the size of the phototubes ($8''$ instead of $10''$
for the actual detector). Each optical module consists of a $17''$
pressure-resistant glass sphere housing an $8''$ Hamamatsu R5912
photomultiplier tube embedded in silicone gel to ensure a good optical
coupling. A cage made from an alloy with a high magnetic permittivity
surrounds each tube, shielding it from the terrestrial magnetic field. To
monitor the efficiency of the photomultiplier tube, an alpha source
was placed in each optical module.

\begin{figure}[h] \centering
  \mbox{\subfigure[Line for data sets 1 and~2.]
  {\epsfig{file=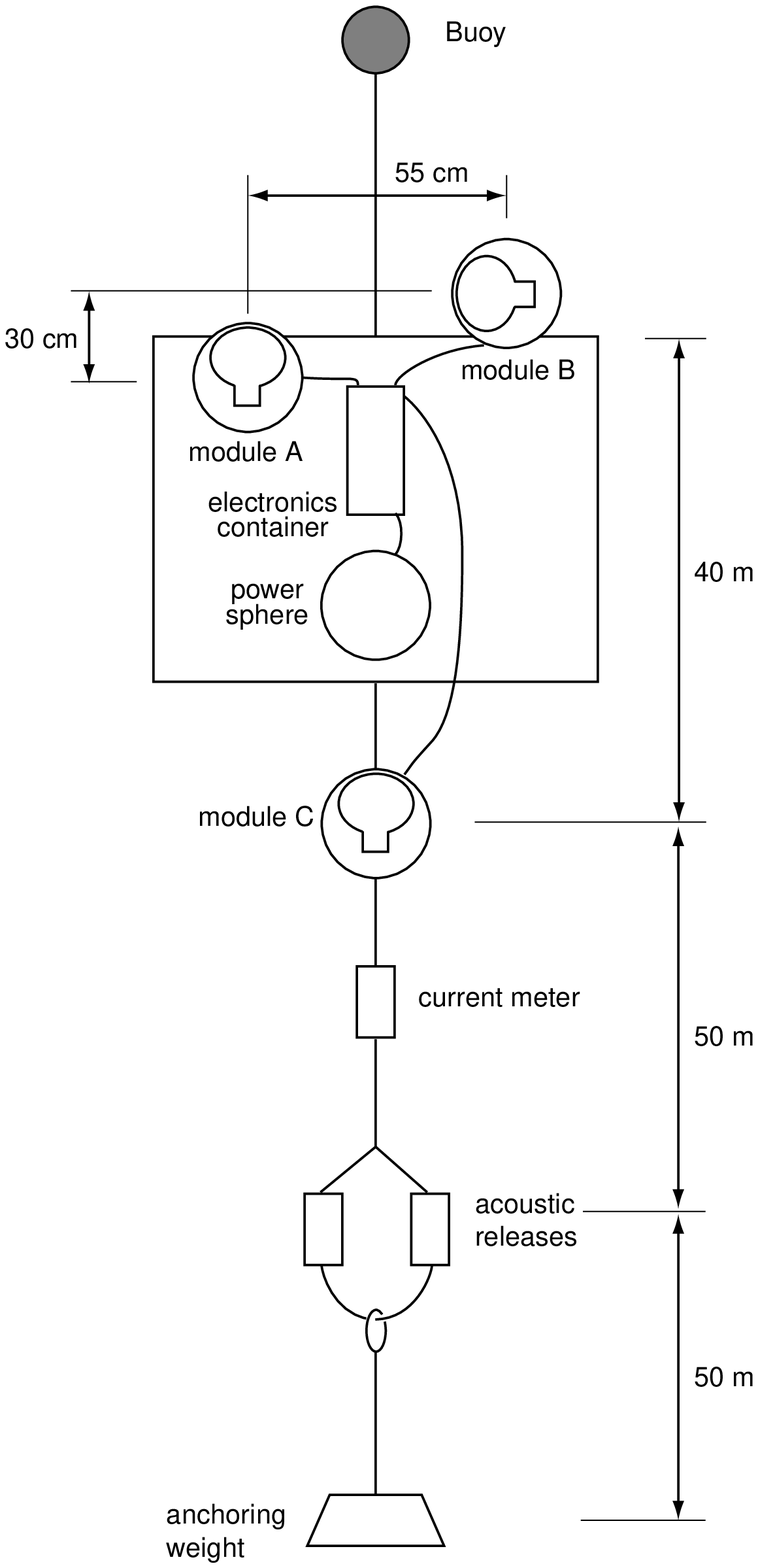,height=9.7cm}} \qquad
  \subfigure[Line for data set 3.]
  {\epsfig{file=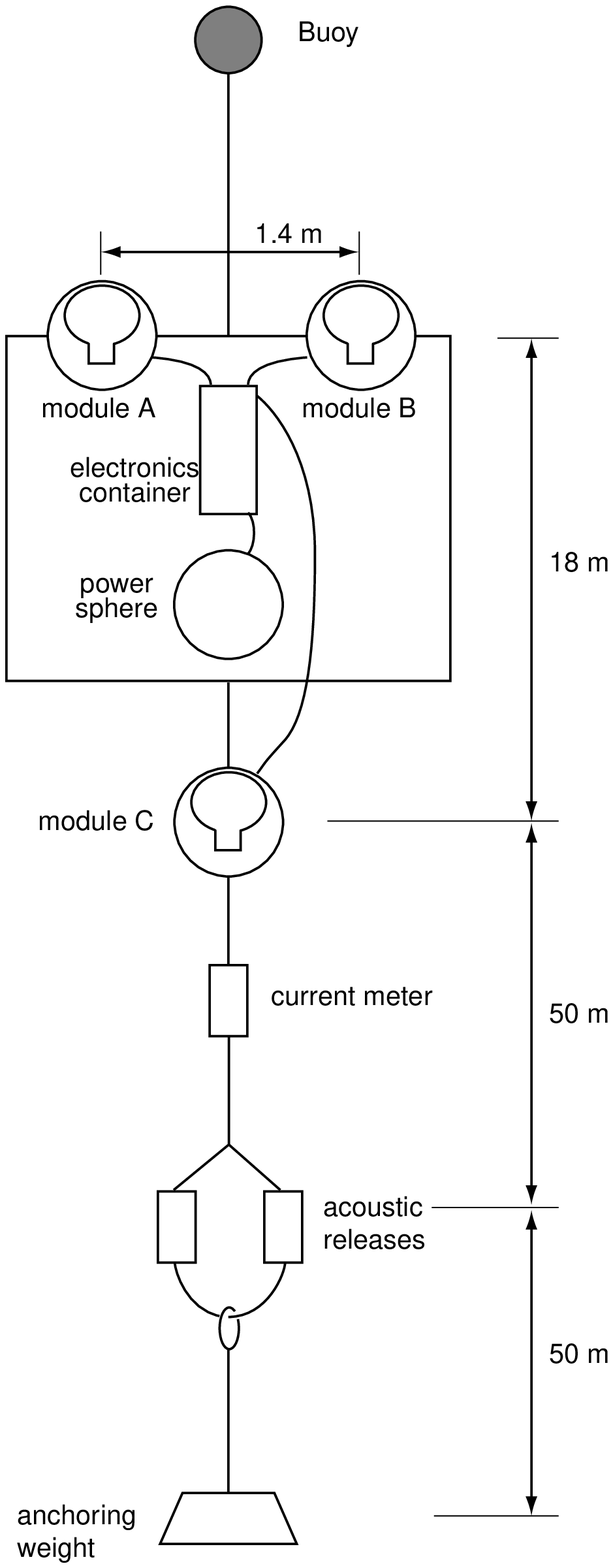,height=9.7cm}}}
  \caption{The two configurations of the mooring lines
  immersed for the measure of the contribution from background
  light. Elements are not to scale.}\label{sch_lines} 
\end{figure}

The optical modules were deployed on ``mooring lines'' in the two
configurations shown in figure~\ref{sch_lines}.  Our standard mooring
line has (as indicated in figure~\ref{sch_lines} from bottom to top)
an anchoring weight (iron chains, about 300~kg) followed by a pair of
releases controlled remotely via an acoustic transducer and mounted in
parallel for redundancy, and a current meter to measure the velocity
of the water past the line. Buoys placed at the top of the line keep
it taut and vertical. An Argos beacon and a flasher beacon assist in
locating the line after release, during recovery. For the measurement
described here, a rigid frame is incorporated into the mooring line,
holding two optical modules (``A'' and ``B''), an anodized aluminum
cylinder housing the electronics and the data acquisition system, and
a ``power'' sphere holding batteries.  A third optical module ``C''
(not present in all tests) is attached to the mooring line, below this
frame. The orientation of the optical modules and the distances
between them were varied as shown in figure~\ref{sch_lines}.

\noindent


\subsection{Data acquisition}

   The electronics and data acquisition system required for our
stand-alone tests have been developed around a MBX~9000 acquisition
card from MII (Micro Informatique pour l'Industrie), equipped with
two~Mbytes of RAM for data storage~\cite{gournay}.  Digital and
analogue I/O's including two serial links are available for
communication with each specific piece of equipment such as current-meter,
acoustic modem and the MBX extension card that holds the electronics
needed for each test.  A Unix-like real-time operating system runs on
the processor.  A configuration file describing the test sequence is
read by the acquisition program at the startup of the processor.
Single rates as well as prompt and delayed coincidences are counted
for two detectors (either A and B, or A and C) on four 100~MHz
scalers. The recorded data comprises:
\begin{itemize}
\item
the time needed to reach a preset count on the singles scaler
for module A, 
\item
the counts reached during this period on the B/C singles scaler,
\item
the coincidence counts (with a 100~ns time gate width) on the AB/AC
coincidence scaler,
\item 
the random coincidence counts with B/C obtained, on the fourth scaler, by
delaying the signals from detector A.
\end{itemize}
With deadtimes not exceeding 1~$\mu$s, this technique gives
sensitivity to frequencies up to a few~MHz. Parameters such as tilt,
current speed, direction of the Earth's magnetic field, temperature,
depth and salinity are monitored during the entire detector operation.


\subsection{Data sets}

The analysis of three sets of data is presented. 
\begin{description}
\item[Set 1] was taken 20~nautical miles off Toulon
(42$^{\circ}$50'~N, 6$^{\circ}$10'~E, called hereafter the ``ANTARES
site'') at a depth of 2430~m, measured at the level of the acoustic
releases, with the configuration shown in
figure~\ref{sch_lines}a. Twelve hours of data were collected during
the night of October~7-8,~1997.

\item[Set 2] was taken with the same configuration but at a site
20~nautical miles off Porto, Corsica (42$^{\circ}$22'~N,
8$^{\circ}$15'~E) at a depth of 2680~m. Data were collected during a 20
minute period, three days a week, from October~10,~1997 to
February~12,~1998.

\item[Set 3] was taken at the ANTARES site at a depth of 2430~m using
the configuration of figure~\ref{sch_lines}b. Data were collected for
4~hours, three days a week from March~10, 1998 to April~8, 1998.

\end{description}

The counting rates were measured successively at two different
amplitude thresholds corresponding to 0.3 and 2 times the average
photoelectron pulse height. The set-up used allows the study of
correlations between signals at distances of 55~cm, 1.4~m, 18~m and
40~m, as well as their dependence on current velocity.


\section{ANALYSIS}

\begin{figure}[h] \begin{center}
\epsfig{file=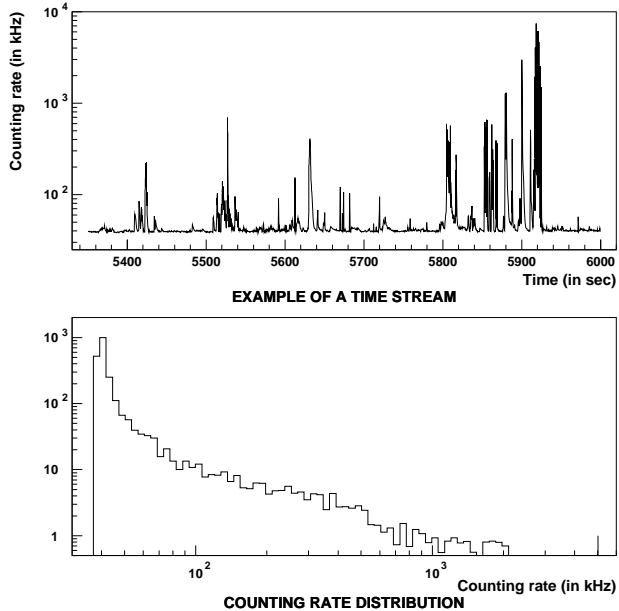,width=.65\textwidth}\caption{Top panel:
typical time dependence of the counting rate at the 0.3~photoelectron
level (from set~1). Bottom panel: distribution of the counting rates
for the above time stream. There is a sharp cutoff at a rate of $\sim
35$~kHz, corresponding to the level of the background light
continuum. The high frequency component of the spectrum corresponds to
activity in the burst regime.}
\label{freq_distrib} 
\end{center} \end{figure}

The data acquired as above were converted into counting rates as a
function of time for each of the optical modules. A typical time
stream is shown in figure~\ref{freq_distrib}, along with the
distribution of the counting rates.

As can be seen in figure 3, the data stream consists of two
components: a continuum base rate of a few tens of kHz, varying slowly
on time-scales of a few hours, and sharp peaks lasting a few seconds
and rising to tens of MHz (``bursts''). These two contributions will
be studied separately.


\subsection{Background light --- continuous component}\label{sec_opt_backgd}

The continuous component of the background light is defined as the
lower envelope of the distribution of the counting rate of a given
optical module as a function of time. The activity fluctuates by as
much as 14~kHz on a time scale of a few hours (see
table~\ref{tab_backgd} and figures~\ref{coinc16} and \ref{coinc17},
top panels). This level varies simultaneously on all optical modules
even when located 40~m apart. No correlation is observed between this
modulation and the current velocity.

\begin{table}[h]
\begin{center}
\caption{Variation of the level of the continuous component 
of the background light. The impact of fouling is negligible compared
to the modulation amplitude.}
\label{tab_backgd}
\begin{tabular}{|c|c|c|c|c|}
\hline
\multirow{2}{1.3cm}{\centering \em Set } & {\em Time of} & 
\multirow{2}{3.3cm}{\centering \em Location} &\multicolumn{2}{c|}{\em Background (in kHz)}\\
\cline{4-5}
 & {\em year} & &{\em Minimum} &{\em Maximum}\\
\hline
1 & Oct. 97 & ANTARES site & 37 & 47\\
2 & Oct. 97 - Feb. 98 & Corsica & 24 & 27 \\
3 & Mar. 98 - Apr. 98 & ANTARES site & 20 & 34\\
\hline
\end{tabular} 
\end{center} \end{table}

A fraction of this continuous component is expected to be caused by
the radioactive decay of potassium. The water salinity
($38.5\,^\circ\!/\!_{\circ\circ}$) is constant in time and the same
at both sites, so the $^{40}\rm K$ contribution must also be constant
in time, unless the water transparency fluctuates by a large
amount. We therefore estimate an upper limit on the $^{40}\rm K$
contribution as the lowest measured rate: 20~kHz.

The modulation in the continuum is probably caused by a variable
bioluminescence component. The amplitude of the modulation is larger
in the ANTARES site than in Corsica, which hints at a higher activity
of the continuous luminescence background, although we cannot exclude
seasonal effects or any long-period variation.

Setting the threshold to 2~photoelectrons suppresses the background
counting rate by about a factor of 100, reducing it to $\sim 300$~Hz,
with a contribution of $\sim 150$~Hz from the $^{40}\rm K$ contained
in the glass sphere and from the emission of the $\alpha$ sources
located in each of the optical modules (used to monitor the PMT
efficiency).

The 2~photoelectron background is well correlated with the 0.3
photoelectron background in the sense that it exhibits a coherent
modulation with the same relative amplitude as the 0.3~photoelectron
signal. This suggests that most of the remaining 2~photoelectron
background comes from the tail of the 1~photoelectron distribution. 


\subsection{Background light --- burst regime}\label{sec::burst}

Short bursts in the counting rate are observed, as shown in
figure~\ref{freq_distrib}, probably due to the passage near the
detector of light emitting organisms. The modulation of the continuous
component is not correlated with periods of high burst activity, so
the two effects are probably caused by distinct populations. This is
observed in each of the three data sets.

The burst rate may be defined as the fraction of time a given optical
module exhibits a counting frequency above 200~kHz at the
0.3~photoelectron level. This threshold is chosen because the data
acquisition system is expected to suffer significant deadtime above
200~kHz. The mean values (i.e. averaged over the entire test period)
of the rates $R_A$, $R_B$ and $R_C$ (for modules A, B and C
respectively) are given in table~\ref{tab_bio}. The rates $R_{AB}$ and
$R_{AC}$ are defined as the fraction of time that the two modules A
and B or A and C are simultaneously exhibiting singles counting rates
above 200~kHz.

\begin{table}[h]
\begin{center}
\caption{Fraction of time a given optical module is affected 
by bursts (see definition in text). The values for 18~m are derived
from a single sequence of data and thus affected by a larger
uncertainty. Separations of 0.55 and 1.4~m are obtained with modules A
and B, while separations of 18 and 40~m are with modules A and C.}
\label{tab_bio}
\begin{tabular}{|c|c|c|c|c|c|}
\hline
{\em Set --- Location} & 
\begin{tabular}{c} {\em Distance}\\{$AB/AC$} \end{tabular}& 
\begin{tabular}{c} {\em Time of}\\{\em year} \end{tabular}&
{$R_A$} & 
{$R_{B/C}$} &
{$R_{AB/AC}$}\\
\hline
{ 1 - ANTARES site} & 0.55 m & Fall &4.8\% & 6.2\% & 4.4\% \\
{ 1 - ANTARES site} & 40 m & Fall &4.6\% & 6.0\% & $< 0.1$\% \\
{ 2 - Corsica} & 0.55 m & Fall & 1.2\% & 3.1\% & 1.0\% \\
{ 3 - ANTARES site} & 1.4 m & Spring &1.6\% & 1.5\% & 1.0\% \\
{ 3 - ANTARES site} & 18 m & Spring & 2.1\% & 1.3\% & $< 0.1$\% \\
\hline
\end{tabular} 
\end{center} \end{table}

\begin{figure}[h] \begin{center} \vspace{-.5cm}
\epsfig{file=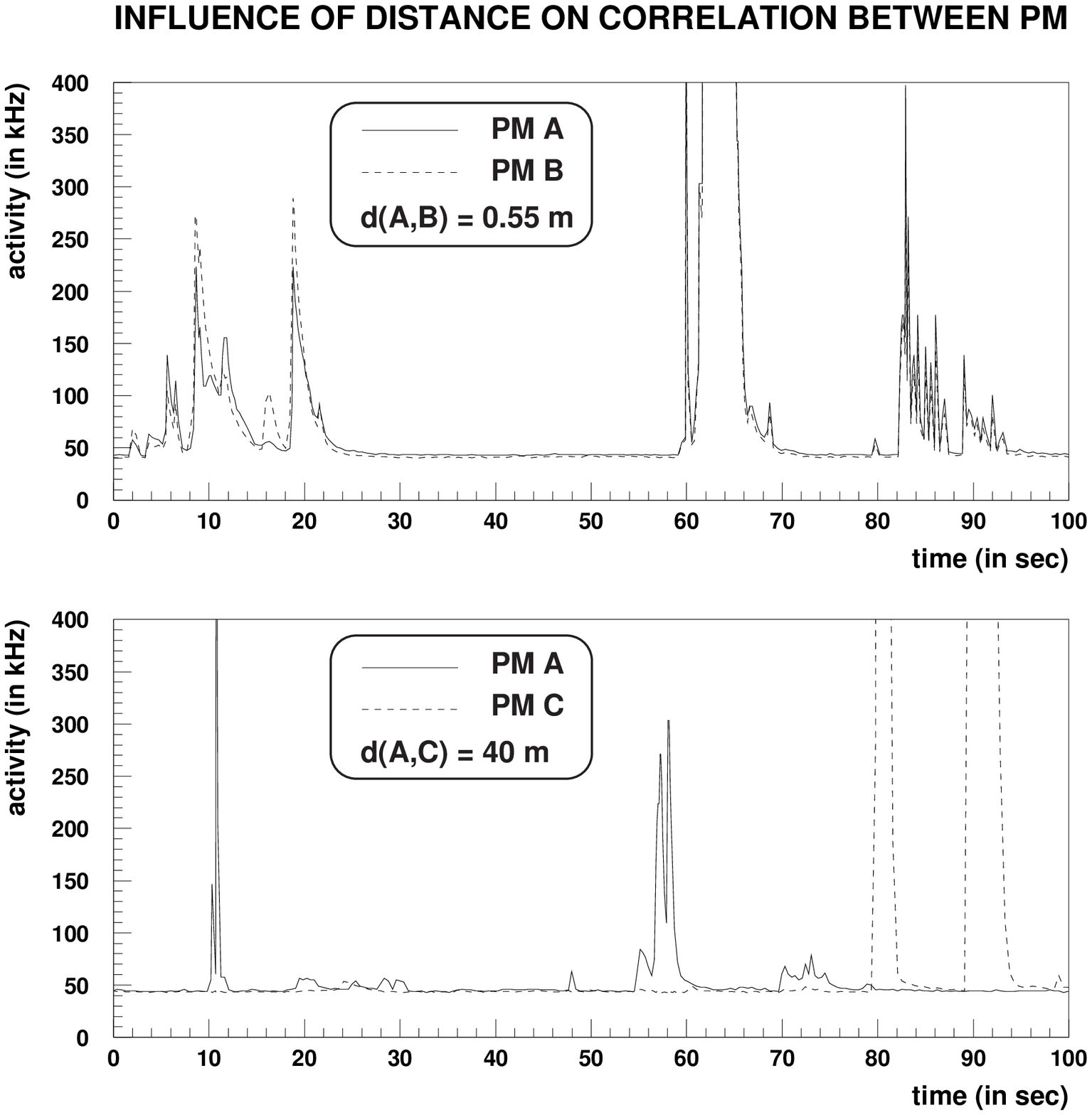,width=.5\textwidth}\caption{Example of time
streams illustrating the strong correlation between the counting
frequencies of modules located 0.55~m apart and the negligible
correlation for modules 40~m apart (data from set~1).}
\label{correl}
\epsfig{file=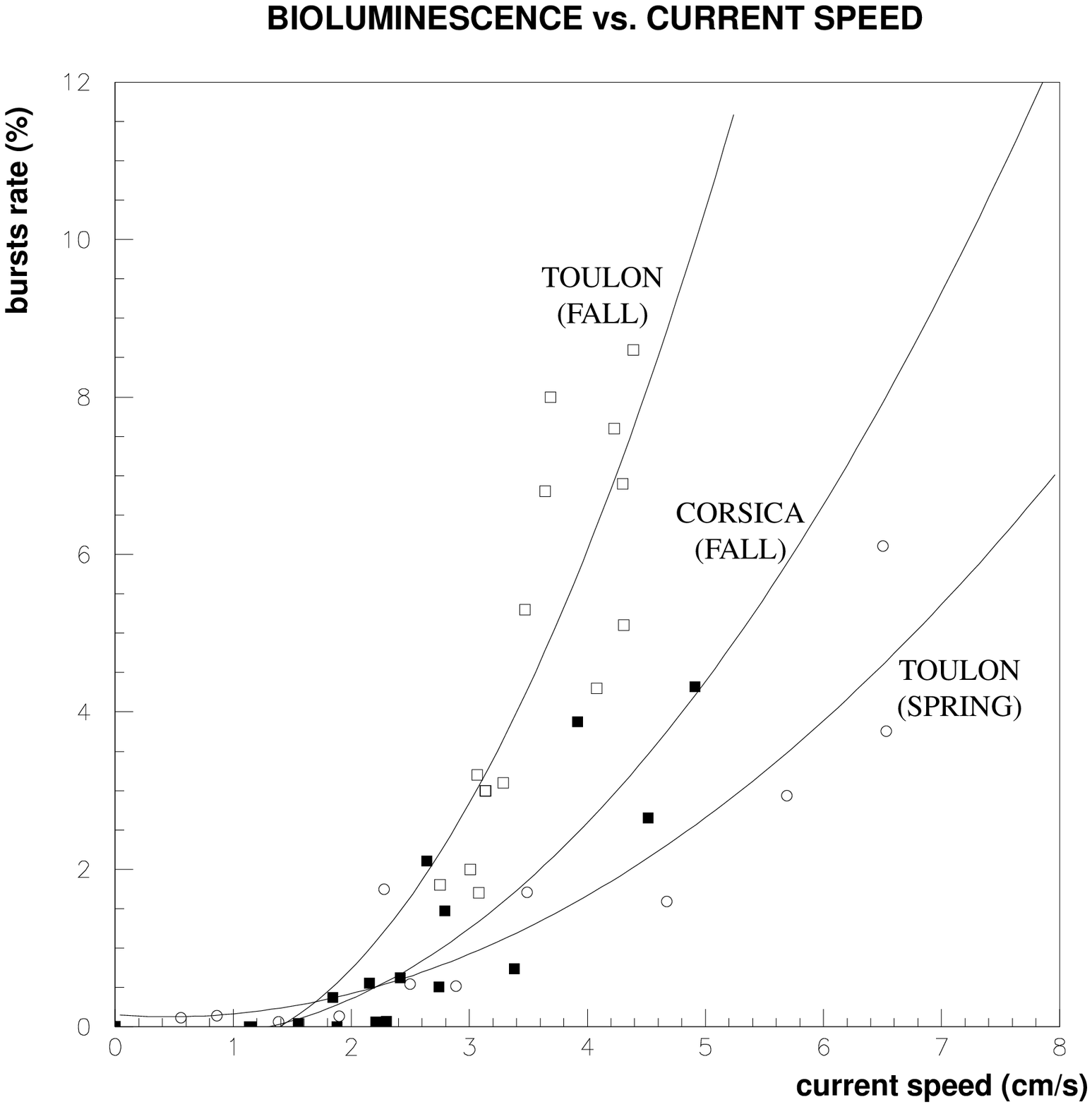,width=.5\textwidth}\caption{Correlation
between the burst rate ($R_{AB}$ as defined in the text) and the
current velocity, for all three tests. A fit with a second degree
polynomial is superimposed on top of the data from each individual
test.}
\label{current}
\end{center} \end{figure}

For data sets 1 and 2, $R_B$ is systematically higher than $R_A$ (A
and B are the two ``close'' modules), which can be understood by the
different orientation of these two modules: A is looking up while B is
looking towards the horizontal (cf. figure~\ref{sch_lines}). As
mentioned below, the bursts are correlated with the current (an
horizontal westwards stream, in the ANTARES site) which therefore
could influence differently the burst rates on modules A and B. This
is corroborated by the fact that in set 3 both modules face the zenith
and detect the same burst rate.

There is a clear indication for correlation between the periods of
burst activity for two modules located near one another, whereas this
correlation is negligible for distant modules: two optical modules
within 1.4~m of each other detect bursts simultaneously, while two
modules located 40~m apart are affected at random uncorrelated times
(rate for both being affected simultaneously smaller than 0.1\%). Two
modules 18~m apart also seem to be affected at random uncorrelated
times, but additional data are required to assess this result with
more confidence. An example of time streams and their burst
correlation is illustrated in figure~\ref{correl}.

The dependence of burst activity on current velocity is emphasized in
figure~\ref{current}. A strong correlation between the two variables
is observed. However, the results from the three data sets differ
significantly.  It is natural to interpret this as reflecting a
dependence on site and season, but more data would be required to
confirm such an interpretation.  During the test off Corsica, the
current speed was on average smaller than in Toulon and a lower burst
activity was observed.  This statement needs to be validated on a
longer time scale.

These data show that bursts are a local phenomenon affecting a given
optical module at most $\sim 6\%$ of the time. A seasonal dependence
might explain the fact that this value is down to $\sim 2\%$ in
set~3.

\subsection{Coincidences between optical modules}

\subsubsection{Coincidence rates}

\begin{figure}[h] \begin{center}
\epsfig{file=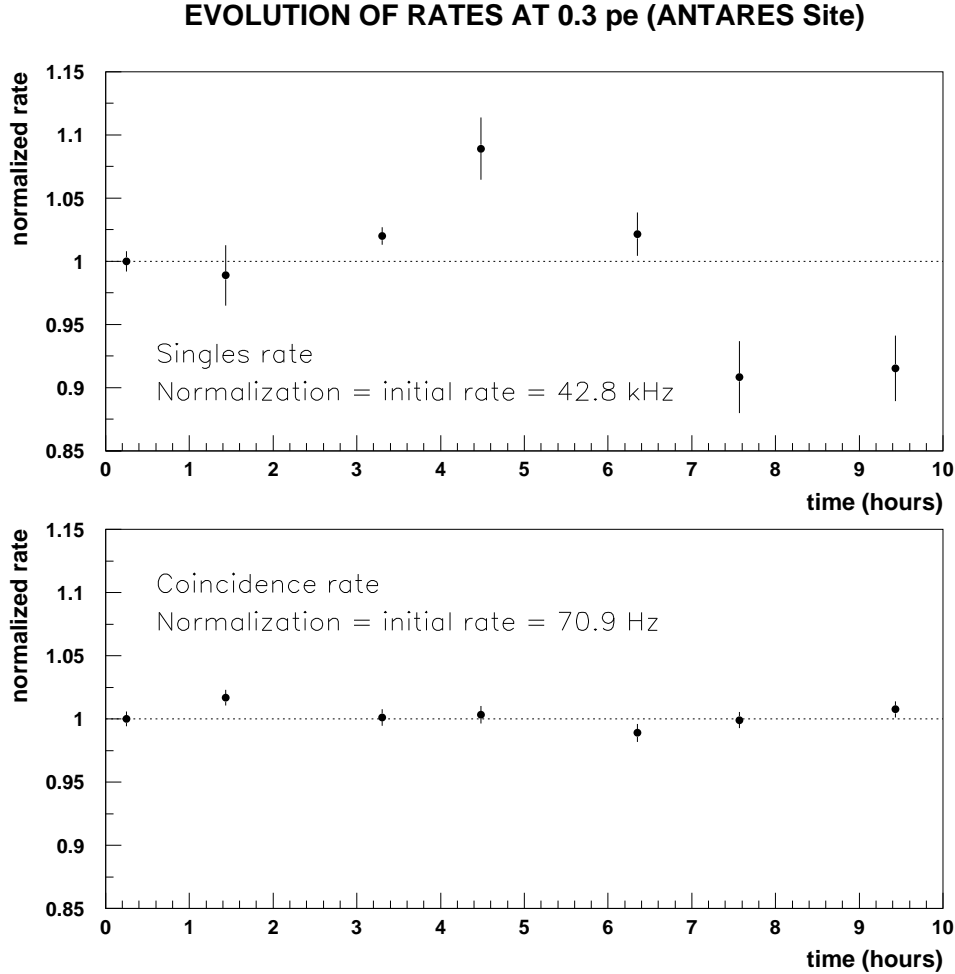,width=\textwidth}\caption{Evolution with
time of the singles rate continuous component (top panel) and the
coincidence rate (bottom panel) between modules A and B, in the
ANTARES site (data set~1). Both rates are normalized to their initial
value.}
\label{coinc16}
\end{center} \end{figure}

Figure \ref{coinc16} illustrates the evolution with time of the
coincidence rates between modules A and B measured at the
0.3~photoelectron level with a 100~ns time window, for data set~1,
where periods of bursts have been cut out. The variation of the
singles rate over a 12~hour period (top panel) is simply the variation
of the continuous component as discussed in
section~\ref{sec_opt_backgd}. In contrast, the coincidence rate is
constant over the same period. This indicates that coincidences are
produced by the $^{40}$K decays, which occur at a constant rate, and
not by the variable component (presumably bioluminescence) of the
continuous contribution.

\begin{figure}[h] \begin{center}
\epsfig{file=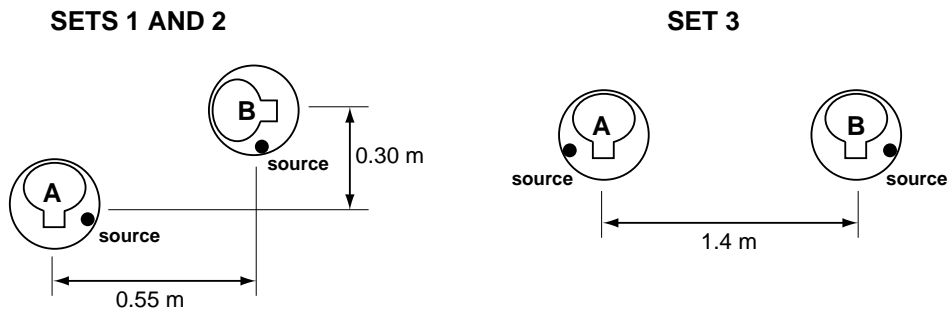,width=.9\textwidth}\caption{Position of the
$\alpha$ sources in the glass spheres.} \label{sources}
\end{center} \end{figure}

The coincidence rates, corrected for accidentals by subtracting from
the measured rate the delayed coincidence rate, are summarized in
table~\ref{tab_coinc} (``Coincident rate''). The contribution to this
rate from the $\alpha$ source contained in each optical module (see
figure~\ref{sources}) was measured in air and is also indicated in the
table. The correction applied is affected by a systematic uncertainty
due to the different media (air vs. water). The reduced coincidence
rates, after subtraction of the $\alpha$ source contribution, are
given in the last column (``Reduced rate''). We observe a null
coincidence rate at the 2~photoelectron level.

\begin{table}[h]
\begin{center}
\caption{Coincidence rates for all three tests (see text for 
details).} 
\label{tab_coinc}
\begin{tabular}{|c|c||c|c||c|}
\hline
{\em Set} & \begin{tabular}{c}{\em Threshold}\\{\em (in pe)} \end{tabular}&
\begin{tabular}{c} {\em Coincident}\\{\em rate (in Hz)} \end{tabular}&
\begin{tabular}{c} {$\alpha$ {\em sources}}\\{\em (in Hz)} \end{tabular} &
\begin{tabular}{c} {\em Reduced}\\{\em rate (in Hz)}\end{tabular}\\
\hline
\hline
1 & 0.3 & 71 & 52 & 19 \\
1 & 2.0  &   15 & 17 & $\sim 0$ \\
\hline
2 & 0.3 & 71  & 52 & 19 \\
\hline
3 & 0.3 & 11 & $\sim 0$& 11 \\
3 & 2.0 & 0 & $\sim 0$& 0 \\
\hline
\end{tabular} 
\end{center} \end{table}

The coincidence rate from $^{40}$K is thus 19~Hz at the 0.3~pe
level for two modules located 55~cm apart, whether near Toulon or near
Corsica, and down to 11~Hz at the 0.3~pe level for two modules located
1.4~m apart, for the geometries shown in figure~\ref{sources}. The
continuum background noise can thus be suppressed by three orders of
magnitude by requiring coincidences, as compared to the singles rate
(see section \ref{sec_opt_backgd}). At the 2~pe level, the coincidence
rate vanishes.

\subsubsection{Comparison with a Monte Carlo calculation}

An estimate of the counting rates due to $^{40}$K was obtained using a
Monte Carlo simulation based on GEANT 3.21~\cite{geant}. A $^{40}$K
activity of 13 Bq per liter and an effective attenuation length of 41
m (as measured at the same site in December 1997~\cite{attenuation})
were assumed (the counting rate is directly proportional to the
effective attenuation length). All the components of the optical
module were taken into account (in terms of efficiency or shielding).

The simulation yields the following rates at the 0.3~photoelectron
threshold:

\begin{center}
\begin{tabular}{|lc|cccc|}
\hline
singles rate: && 18 & $\pm$ & 3 & kHz \\
\hline
\multirow{2}{3.5cm}{coincidence rates:} & Sets 1 \& 2 & 28 & $\pm$ & 4 & Hz \\
& Set 3 & 11 & $\pm$ & 3& Hz \\
\hline
\end{tabular}
\end{center}
Given the uncertainty on the $\alpha$ source correction, these numbers
are in good agreement with the observations described previously. As
observed in the data, the simulation gives negligible coincidence
rates at the 2~photoelectron threshold (less than 1~Hz).

\subsubsection{Effect of fouling}

\begin{figure}[h] \begin{center}
\epsfig{file=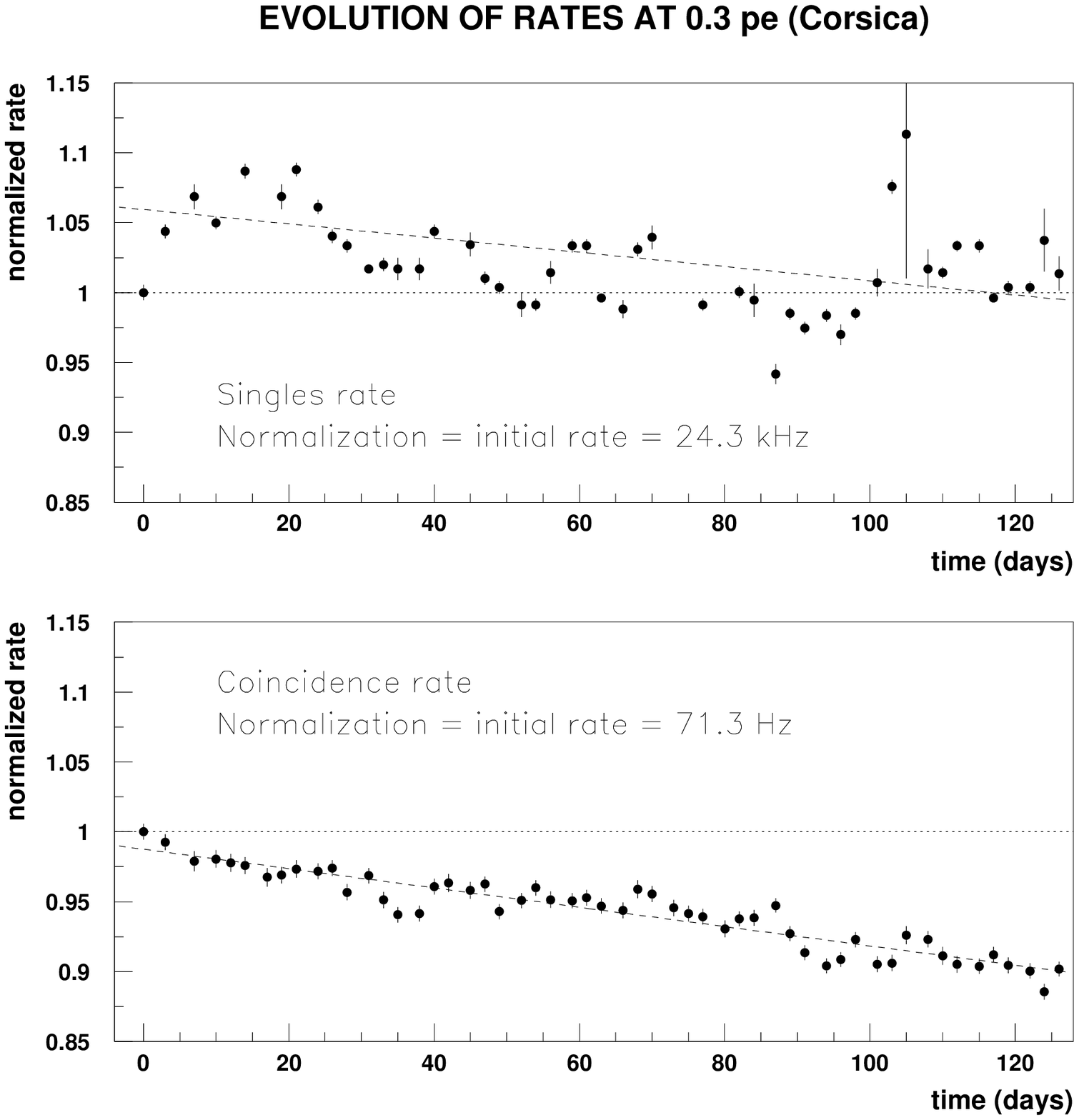,width=\textwidth}\caption{Evolution with
time of the singles rate continuous component (top panel), and the
coincidence rate (bottom panel) between modules A and B, in Corsica
(data set~2). Both rates are normalized to their initial value.}
\label{coinc17}
\end{center} \end{figure}

The decrease of the coincidence rate with time is due to fouling on
the modules (cf figure~\ref{coinc17}, bottom panel), whether due to
sedimentation or to the growth of bacteria on the surfaces. It is
mostly noticeable on the evolution of the coincidence rate since the
evolution of the singles rate is dominated by the large modulation of
the background light continuum. On a large time scale however, as for
the test in Corsica, both singles and coincidence rates decrease in a
similar fashion (figure~\ref{coinc17}). No effect is noticeable in the
first test on the ANTARES site, which only lasted 12 hours
(figure~\ref{coinc16}, bottom panel).

The global effect of fouling can be estimated from the decrease in the
coincidence rates observed in sets~2 (120 days near Corsica) and 3 (30
days near Toulon), taking into account the presence of the $\alpha$
sources (see figure~\ref{sources}). Let $t_A$ and $t_B$ be the local
transmission efficiencies near the $\alpha$ sources of modules A and B
(for the light emitting module), $<\!t\!>$ be the transmission
efficiency averaged over the optical module (for the light receiving
module), $f$ the coincidence rate detected, $f_\alpha$ the rate of the
$\alpha$ sources (subscripted by A and B in the following equation for
the sake of clarity, but assumed to be equal) and $f_{^{40}{\rm K}}$
the coincidence rate generated by the $^{40}{\rm K}$ background. In
general, we can write
\begin{equation}
f = <\!t_A\!><\!t_B\!>f_{^{40}{\rm K}} +
    <\!t_A\!>t_B f_{\alpha,\,B} + 
    <\!t_B\!>t_A f_{\alpha,\,A} \label{eq_fouling}
\end{equation}

In the second and third terms of the right-hand side of this equation,
the $<\!t\!>$ coefficients should be affected by a second-order
correction since regions of the detector facing the $\alpha$ source
contained in the other sphere are actually favored. This effect is not
taken into account.

Equation \ref{eq_fouling} can be simplified and solved to first order
for the most significantly affected module(s). An extensive study of
the effect of fouling has been performed by the ANTARES collaboration
with a dedicated set-up~\cite{fouling}, showing that most fouling is
for surfaces pointing directly at the zenith, decreasing quickly with
angle.  The local transmission efficiencies $t_A$ and $t_B$ are thus
assumed to be constant since the $\alpha$ sources are located below
the equator of the glass sphere. The contributions from $^{40}{\rm K}$
and the $\alpha$ sources are also assumed constant. For set~2,
$<\!t_B\!>$ is assumed constant since the module is pointing at
$90^\circ$ with the respect to the zenith, while in set~3, we consider
that $<\!t_A\!> = <\!t_B\!>$ (both facing the zenith). The effect of
fouling on a module in the ANTARES site is then estimated to be:
\begin{equation}
 <\!t\!>_{(\theta = 0^\circ,\; \Delta t = 30{\rm \,days})} = 76\%
\label{fouling_toulon}\end{equation}
where $\theta$ is the angle between the axis of the module and the
zenith and $\Delta t$ the duration of the experiment. For the
test immersed in Corsica, we measure:
\begin{equation}
 <\!t\!>_{(\theta = 0^\circ,\; \Delta t = 120{\rm \, days})} = 87\%
\label{fouling_corse} \end{equation}
The Corsica site, in fall, thus seems cleaner than that near
Toulon. It could, however, be only a seasonal fluctuation.

Significant residual fluctuations can be observed around the average
decrease of the coincidence rate (see plot of figure~\ref{coinc17},
bottom panel). They are probably caused by a non-constant fouling
rate, due for instance to the temporary ``cleaning'' effect of
current. On a 30~day period, these fluctuations can induce an
uncertainty on the rate of fouling which can account for the
difference in the estimates obtained in equations~\ref{fouling_toulon}
and \ref{fouling_corse}.


\section{DESIGN CONSTRAINTS}

The ANTARES collaboration proposes to build a first detector
consisting of 1000~optical modules housing $10''$ photomultiplier
tubes (instead of the $8''$ tubes used for the measurements presented
in this paper --- singles rates given in this article have therefore
been scaled by 1.6 for the following estimates), leading to an
effective area of 0.1~${\rm km}^2$. We here infer the constraints put
on the design of an undersea neutrino telescope from the measured
characteristics of the background light.

\subsection{Trigger and electronics}
In order to cope with the background light and the fouling, we plan to
arrange optical modules in clusters of 3 optical modules facing away
from each other with their axis making an angle of $45^\circ$ with the
vertical, and to use twofold local coincidences in the trigger. This
reduces the accidental rate per cluster to a few 100~Hz. The rate due
to $^{40}$K coincidences in the planned configuration is negligible. A
majority trigger condition for the whole detector is built from these
coincidences. Requiring four coincidences yields a trigger rate of a
few kHz, provided that clusters affected by bursts of bioluminescence
are inhibited.

An analogue memory in an ASIC has been designed in which the
photomultiplier pulses and arrival times can be stored. The memory
depth is sufficient to cover the time needed for the trigger to be
built and a read-out request to be sent back to the whole
array~\cite{ARS}. Given the size of the detector, this time comes
mostly from the propagation of signals to and from the central node
(10~$\mu$s). If a request is received during a time window constrained
by the trigger causality, the data are digitized and sent to the shore
station. They are otherwise discarded.

The deadtime induced by the counting rate will be negligible up to
several 100~kHz. When integrated over the distribution shown in
figure~\ref{freq_distrib} (bottom) and the spatial distribution
inferred from table~\ref{tab_bio}, the overall effect of this deadtime
is less than a ~5\% inefficiency distributed randomly over the
detector.

\subsection{Muon track determination}

The background light has two effects : below the maximum rate which
can be handled by the electronics, it results in spurious hits added
to those coming from the muon track; above this rate it will
temporarily blind parts of the detector, as mentioned above.  These
blind spots can be located as a function of time, and their effect on
the acceptance inferred, by monitoring the counting rate of all
optical modules.

Monte Carlo studies show that for a majority trigger requiring at
least four coincidences, each event contains about ten additional
single photomultiplier tube hits coming from the Cherenkov light
emitted by the muon.

A 50~kHz background counting rate produces about 100~isolated noise
hits randomly distributed over the entire volume, during the 2~$\mu$s
muon time of flight. A first-try track fit is obtained by fitting a
straight line through the hits on coincident optical module pairs.
Limiting the useful volume to a 50~m radius cylinder around this
initial track, and the time difference with respect to this
preliminary fit prediction to 200~ns, only 1~noise hit remains on
average. This causes a negligible distortion to the calculated muon
direction.


\section{CONCLUSION}

The counting rates due to background light have been measured at two
different sites. It is composed mainly of three independent
components: a constant rate due to the beta decay of $^{40}$K
naturally present in the sea salt, another slowly fluctuating on time
scales of a few hours, and one exhibiting bursts a few seconds long,
probably due to the luminescence of living organisms.

We observed a strong correlation between the undersea current velocity
and the bursts, although the absolute dependence seems to be affected
by seasonal variations. While the bursts occur simultaneously for two
optical modules located less than 1.4~m apart, they are uncorrelated
for modules located more than 40~m apart. The modulation of the
continuous component, however, is correlated on at least a 40~m range.

Measured coincidence rates are in agreement with a Monte Carlo
simulation of the Cherenkov light caused by the beta decay of
$^{40}$K. Long term decreases of these coincidence rates are
compatible with independent measurements of optical fouling.

The background light has only a minor effect on the performance of an
undersea neutrino telescope. An electronics system is being developed
which can handle rates up to several 100~kHz without significant deadtime.
A tight timing window reduces background hits to a rate that does not
affect the determination of the muon direction and energy.


\begin{thebibliography}{99}
\bibitem{proposal} 
ANTARES proposal, http://antares.in2p3.fr/antares/proposal99.html. 

\bibitem{markov} Markov, M.A., in Proc. of Annual International Conference of High Energy Physics, Rochester (1960).

\bibitem{gournay} Gournay, J.F. et al.,
ICALEPCS 97 proceedings, Beijing (1997).

\bibitem{geant} Geant3.21 manual, 
http://wwwinfo.cern.ch/asdoc/geant\_html3/geantall.html, PHYS325.

\bibitem{attenuation} 
Palanque-Delabrouille, N., ICRC 99 proceedings, Salt Lake City (1999).

\bibitem{fouling} de Botton, N., ICRC 97 proceedings, Durban (1997).\\
Palanque-Delabrouille, N., ICRC 99 proceedings, Salt Lake City (1999).

\bibitem{ARS} 
Lachartre, D., Beaune 99 proceedings, Beaune (1999).
\end{thebibliography}
\end{document}